\documentclass[12pt,a4paper]{article}
\usepackage[T1]{fontenc}
\usepackage{amsmath,amssymb}
\usepackage{amsthm}
\usepackage{verbatim}
\usepackage{mathabx}
\usepackage[breaklinks]{hyperref}
\usepackage{color}
\usepackage{bm}
\usepackage{graphicx}
\usepackage{multirow}

\newtheorem{lemma}{Lemma}
\newtheorem{theorem}{Theorem}
\newtheorem{corollary}{Corollary}

\theoremstyle{definition}
\newtheorem{definition}{Definition}

\newtheorem{oldtheorem}{Theorem}

\newcommand{\ifdraft}[1]{}
\definecolor{aocolour}{rgb}{0.7,0.8,1}
\definecolor{spcolour}{rgb}{1,0.4,0.4}

\newcommand{\rank}{\mathop{\mathrm{rank}}}

\newcommand{\boundfunc}[1]{f(#1)}

\DeclareMathOperator\End{End}

\renewcommand{\epsilon}{\varepsilon}

\begin{document}

\sloppy

\title{On the rank of the communication matrix for deterministic two-way finite automata}
\author{Semyon Petrov\thanks{Department of Mathematics and Computer Science,
	St.~Petersburg State University, Russia, \texttt{semenuska2010@yandex.ru}.} \and
Fedor Petrov\thanks{Department of Mathematics and Computer Science,
	St.~Petersburg State University, Russia, \texttt{fedyapetrov@gmail.com}.} \and
Alexander Okhotin\thanks{Department of Mathematics and Computer Science,
	St.~Petersburg State University, Russia, \texttt{alexander.okhotin@spbu.ru}.}}

\maketitle

\begin{abstract}
The communication matrix for two-way deterministic finite automata (2DFA) with $n$ states
is defined for an automaton over a full alphabet of all $(2n+1)^n$ possible symbols:
its rows and columns are indexed by strings,
and the entry $(u, v)$ is $1$ if $uv$ is accepted by the automaton,
and $0$ otherwise.
With duplicate rows and columns removed, this is a square matrix of order $n(n^n-(n-1)^n)+1$,
and its rank is known to be a lower bound
on the number of states necessary to transform an $n$-state 2DFA
to a one-way unambiguous finite automaton (UFA).
This paper determines this rank,
showing that it is exactly
$\boundfunc{n}=\sum_{k=1}^n \binom{n}{k-1} \binom{n}{k} \binom{2k-2}{k-1} =(1+o(1)) \frac{3\sqrt{3}}{8\pi n} 9^n$,
and this function becomes the new lower bound
on the state complexity of the 2DFA to UFA transformation,
thus improving a recent lower bound by S. Petrov and Okhotin
(\href{https://doi.org/10.1007/978-3-030-68195-1_7}
{``On the transformation of two-way deterministic finite automata to unambiguous finite automata''},
\emph{Inf.\ Comput.}, 2023).
The key element of the proof is determining the rank of a $k! \times k!$ submatrix,
with its rows and columns indexed by permutations,
where the entry $(\pi, \sigma)$ is $1$ if $\sigma \circ \pi$ is a cycle of length $k$,
and 0 otherwise;
using the methods of group representation theory
it is shown that its rank is exactly $\binom{2k-2}{k-1}$,
and this implies the above formula for $\boundfunc{n}$.

\end{abstract}

\section{Introduction}

A recurring question in the theory of finite automata
is proving lower bounds on the size of automata of different kind.
Several widely used lower bound methods involve a concept
adapted from the field of communication complexity:
the \emph{communication matrix}.
The communication matrix for a formal language $L$ over an alphabet $\Sigma$
is a $\{0, 1\}$-valued matrix
with both its rows and columns indexed by strings over $\Sigma$:
its entry $(u, v)$ contains $1$ if and only if $uv \in L$;
informally, this matrix represents
the information transfer between the prefix $u$ and the suffix $v$
necessary to recognize the membership of their concatenation in $L$.
In general, the matrix is defined with infinitely many rows and columns
corresponding to all strings;
but for a regular language, there are only finitely many distinct rows and columns,
and hence the matrix for $L$ can be regarded as finite.

The usual applications of the communication matrix to estimating the size of automata are as follows.
The number of distinct rows in this matrix
is the size of the minimal deterministic finite automaton (DFA) for that language:
this is the Myhill--Nerode theorem.
A standard method for proving lower bounds on the size of nondeterministic finite automata (NFA)
is the \emph{fooling set} method,
which originates from the work of Aho et al.~\cite{AhoUllmanYannakakis} on communication complexity.
This method was first applied to NFA
by Birget~\cite{Birget} and by Glaister and Shallit~\cite{GlaisterShallit},
and it is based on the size of the largest submatrix of the communication matrix
that does not have certain forbidden patterns.
For unambiguous finite automata (UFA),
that is, for NFA with at most one accepting computation on each string,
the standard lower bound method is \emph{Schmidt's theorem}
(see Schmidt~\cite{Schmidt} for the original proof
and Leung~\cite{Leung} for a modern presentation);
this theorem states that the rank of the communication matrix for a language
is a lower bound on the size of every UFA recognizing that language.
Other lower bound methods for finite automata
employing ideas from communication complexity
were presented by \v{D}uri\v{s} et al.~\cite{DurisHromkovicRolimSchnitger}
and by Hromkovic et al.~\cite{HromkovicSeibertKarhumakiKlauckSchnitger}.

A standard question studied in the literature
is the number of states in finite automata of one kind
needed to simulate an $n$-state finite automaton of another kind.
It remains unknown whether
nondeterministic two-way finite automata (2NFA) with $n$ states
can be simulated by deterministic two-way finite automata (2DFA) with polynomially many states,
and this question is known to be connected to the \emph{L vs.\ NL} problem in the complexity theory,
see Kapoutsis~\cite{Kapoutsis_logspace}
and Kapoutsis and Pighizzini~\cite{KapoutsisPighizzini_logspace}.
Much progress has been made on the complexity of simulating two-way automata by one-way automata.
The possibility of such a transformation
is one of the earliest results of automata theory~\cite{Shepherdson},
and the exact number of states sufficient and necessary to simulate 2NFA and 2DFA
by NFA and DFA
was later determined by Kapoutsis~\cite{Kapoutsis,Kapoutsis_thesis},
who proved that an $n$-state 2DFA can be transformed to a DFA with $n(n^n - (n-1)^n)+1$ states,
and to an NFA with $\binom{2n}{n+1}$ states.
At the same time, Kapoutsis~\cite{Kapoutsis,Kapoutsis_thesis} proved both bounds to be exact,
in the sense that there are $n$-state 2DFA that require that many states to make them one-way.
Other related results include the complexity of transforming two-way automata to one-way automata
recognizing the complement of the original language,
first investigated by Vardi~\cite{Vardi}, followed by Birget~\cite{Birget_state_compressibility},
and later extended to alternating automata by Geffert and Okhotin~\cite{GeffertOkhotin_2afa}.
Attention was given to transforming two-way automata to one-way
in the case of alphabets of limited size:
in the unary case, the complexity was studied by Chrobak~\cite{Chrobak},
Mereghetti and Pighizzini~\cite{MereghettiPighizzini2001},
Geffert et al.~\cite{GeffertMereghettiPighizzini2003,GeffertMereghettiPighizzini}
and Kunc and Okhotin~\cite{TwoWayDFAs},
and later Geffert and Okhotin~\cite{GeffertOkhotin_2dfa_to_dfa}
improved the bounds for alphabets of subexponential size.
Recently, the complexity of transforming sweeping permutation automata to one-way permutation automata
was determined by Radionova and Okhotin~\cite{RadionovaOkhotin}.

For 2DFA of the general form,
their optimal transformations to DFA and to NFA
established by Kapoutsis~\cite{Kapoutsis,Kapoutsis_thesis}
are naturally represented in terms of the following matrix.
For each $n$, there exists a single hardest $n$-state 2DFA
defined over an alphabet with $(2n+1)^n$ symbols,
and the communication matrix for this 2DFA
contains the communication matrices for all languages
recognized by any $n$-state 2DFA as submatrices.
This matrix deserves to be called \emph{the communication matrix for $n$-state 2DFA},
and the results of Kapoutsis imply
that the number of unique rows in this matrix is exactly $n(n^n-(n-1)^n)+1$,
whereas the maximum size of the fooling set in this matrix is exactly $\binom{2n}{n+1}$.
Another fact, a consequence of Schmidt's theorem, is that the rank of this matrix
is a lower bound on the complexity of transforming 2DFA to UFA.

State complexity of UFA is an active research topic.
Leung~\cite{Leung1998,Leung} studied relative succinctness of NFA with different degrees of ambiguity,
and, in particular, proved
that transforming an $n$-state UFA to a DFA requires $2^n$ states in the worst case~\cite{Leung}.
For the case of a unary alphabet, Okhotin~\cite{ufa_sc} showed
that the state complexity of transforming an $n$-state UFA to DFA
is of the order $e^{\Theta((\ln n)^{2/3})}$,
which was recently sharpened to $e^{(1+o(1))4^{1/3}(\ln n)^{2/3}}$ by F. Petrov~\cite{FPetrov_ufa}.
A notable fact is that the complement of every $n$-state UFA
can be recognized by a UFA with much fewer than $2^n$ states:
this was first proved by Jir\'asek Jr.\ et al.~\cite{JirasekjrJiraskovaSebej},
who demonstrated that at most $2^{0.79n}$ states are sufficient,
and soon thereafter Indzhev and Kiefer~\cite{IndzhevKiefer}
improved the upper bound to $\sqrt{n+1} \cdot 2^{n/2}$.
Lower bounds on the number of states necessary to represent the complement
have been researched as well:
in the case of a unary alphabet,
after an early lower bound $n^{2-\epsilon}$ given by Okhotin~\cite{ufa_sc},
Raskin~\cite{Raskin} discovered the state-of-the-art lower bound of $n^{(\log \log \log n)^{\Omega(1)}}$ states.
Later, G\"o\"os, Kiefer and Yuan~\cite{GoosKieferYuan}
used new methods based on communication complexity
to establish a higher lower bound of $n^{\Omega(\log n)/(\log\log n)^{O(1)}}$ states
for a two-symbol alphabet.
Most recently, Czerwi\'nski et al.~\cite{CzerwinskiDebskiGogaszHoiJainSkrzypczakStephanTan}
carried out a detailed analysis of descriptional and computational complexity of unary UFA,
and, in particular,
showed that every $n$-state unary UFA can be complemented using at most $n^{\log n+O(1)}$ states.

Simulation of 2DFA by UFA was recently investigated 
by S. Petrov and Okhotin~\cite{PetrovOkhotin_2dfa_to_1ufa},
who showed that $n! \cdot 2^n$ states are sufficient in a UFA
to simulate an $n$-state 2DFA,
and also established a lower bound
of $\Omega(\frac{1}{n}(\sqrt{2}+1)^{2n}) = \Omega(5.828^n)$ states
on the complexity of this transformation~\cite{PetrovOkhotin_2dfa_to_1ufa}.
This was actually a lower bound on the rank of the communication matrix for $n$-state 2DFA,
and it was obtained as follows.
First, a series of rank-preserving linear transformations were applied to this matrix,
ultimately reducing it to a diagonal combination of matrices of a simpler form,
which have a succinct combinatorial definition that no longer relies on automata.
Then, rough lower bounds on the rank of the latter matrices were given,
and their sum was taken as the final lower bound on the rank of the communication matrix.

The goal of this paper is to reinvestigate the communication matrix for $n$-state 2DFA,
and to determine its rank precisely.
This, in particular, will lead to an improved lower bound
on the transformation from 2DFA to UFA.

Section~\ref{section_matrix_for_permutations} presents the formula
of S. Petrov and Okhotin~\cite{PetrovOkhotin_2dfa_to_1ufa}
that expresses the rank of the communication matrix for $n$-state 2DFA
in terms of ranks of certain submatrices $P^{(k)}$ of size $k! \times k!$.
The submatrix $P^{(k)}$ has a clear combinatorial definition:
it is a $k! \times k!$ matrix,
with its rows and columns indexed by permutations,
where the entry corresponding to a pair of permutations $(\pi, \sigma)$
is $1$ if $\sigma \circ \pi$ is a cyclic permutation,
and 0 otherwise.
All that is known so far
is that the rank of each $P^{(k)}$ is at least $2^{k-1}$,
but calculations for small $k$ indicate that it is somewhat higher~\cite{PetrovOkhotin_2dfa_to_1ufa}.
The task is to determine its rank precisely.

For uniformity of notation with the mathematical tools used in this paper,
the matrix in question shall be denoted by $P^{(n)}$.
The first step in the calculation of the rank of $P^{(n)}$
is taken in Section~\ref{section_group_algebra},
where the matrix $P^{(n)}$ is represented
as a linear operator $\mathcal{L}_n$
in the group algebra of the symmetric group $S_n$ over the field of complex numbers,
denoted by $\mathbb{C}[S_n]$.
The image of this operator is a linear space,
and its dimension, called \emph{the rank of $\mathcal{L}_n$},
turns out to be the same as the rank of $P^{(n)}$.

The dimension of this linear space is determined using the methods of group representation theory.
In Section~\ref{section_modules},
once the basic definitions are briefly recalled,
the group algebra $\mathbb{C}[S_n]$
is regarded as an $S_n$-module,
that is, a vector space with a linear mapping associated with each permutation in $S_n$.
Then, by the classical results on group representations,
this $S_n$-module is expressed as a direct sum of irreducible $S_n$-modules
(those with no non-trivial proper subspaces),
and hence its dimension is the sum of dimensions of those irreducible modules.
The next Section~\ref{section_specht} recalls
the classical association of Young diagrams to irreducible $S_n$-modules,
so that the number of standard Young tableaux for a diagram
is exactly the dimension of the corresponding module.
Finally, Section~\ref{section_characters} establishes that
for each irreducible $S_n$-module participating
in the direct sum representation $\mathbb{C}[S_n]$,
the restriction of $\mathcal{L}_n$ to this $S_n$-module
has full rank if the corresponding Young diagram is of a hook shape,
and is constant zero otherwise.
This leads to a direct formula for the desired rank of $\mathcal{L}_n$,
which is $\binom{2n-2}{n-1}$.

Thus, the rank of the matrix $P^{(k)}$ is $\binom{2k-2}{k-1}$,
and substituting it into the formula
from the earlier paper by S. Petrov and Okhotin~\cite{PetrovOkhotin_2dfa_to_1ufa}
shows that the communication matrix for $n$-state 2DFA
has rank $\boundfunc{n}=\sum_{k=1}^n {n \choose{k-1}} {n \choose k} \binom{2k-2}{k-1}$.
This becomes the new lower bound on the state complexity of transforming 2DFA to UFA.
In the last Section~\ref{section_asymptotics},
its asymptotics is estimated as $\boundfunc{n}=(1+o(1)) \frac{3\sqrt{3}}{8\pi n}\cdot 9^n$.

\section{The matrix for permutations}\label{section_matrix_for_permutations}

\begin{definition}
Let $L$ be a language over an alphabet $\Sigma$.
The communication matrix for $L$
is an infinite matrix
with both its rows and columns indexed by strings over $\Sigma$.
Its entry $(u, v)$ contains $1$ if $uv \in L$,
and $0$ if $uv \notin L$.
\end{definition}

The communication matrix for a regular language
always has finitely many different rows
and finitely many different columns,
and accordingly can be reduced to a finite matrix by removing duplicate rows and columns.

The next goal is to define, for each $n \geqslant 1$,
a single communication matrix
applicable to all languages recognized by deterministic two-way finite automata (2DFA) with $n$ states.
This is done using the well-known \emph{universal $n$-state 2DFA $A_n$},
defined over an alphabet of size ca.~$n^{2n+1}$ symbols,
so that every $n$-state 2DFA could be obtained from $A_n$
by restricting its alphabet.

\begin{definition}\label{comm_matrix_for_n_state_2dfa_definition}
For each $n \geqslant 1$,
the communication matrix for $n$-state 2DFA
is the communication matrix for the language $L(A_n)$,
with duplicate rows and columns removed.
\end{definition}

Then, the communication matrix for any language recognized by an $n$-state 2DFA
is a submatrix of the matrix in Definition~\ref{comm_matrix_for_n_state_2dfa_definition}.

One of the reasons for determining the rank of the communication matrix for $n$-state 2DFA
is that, according to the following theorem,
it gives a lower bound on the complexity of transforming 2DFA to UFA.

\begin{oldtheorem}[Schmidt~\cite{Schmidt}, see also Leung~\cite{Leung}]\label{Schmidt_theorem}
Let $L$ be a regular language,
and let $(x_1, y_1)$, \ldots, $(x_n, y_n)$ be pairs of strings.
Let $M$ be an integer matrix
defined by $M_{i,j} = 1$, if $x_iy_j \in L$,
and $M_{i,j} = 0$ otherwise.
Then, every UFA for $L$ has at least $\rank M$ states.
\end{oldtheorem}

In the recent paper by S. Petrov and Okhotin~\cite{PetrovOkhotin_2dfa_to_1ufa},
the rank of this communication matrix
was expressed through the rank of the following simpler matrices.

\begin{definition}[{\cite[Lemma~14]{PetrovOkhotin_2dfa_to_1ufa}}]
\label{P_definition}
For each $k \geqslant 1$, the $k! \times k!$ matrix $P^{(k)}$
is defined by $P^{(k)}_{\pi, \sigma} = 1$
if
the permutation $\sigma \circ \pi$ is cyclic,
and $P^{(k)}_{\pi, \sigma} = 0$ otherwise.
\end{definition}

\begin{figure}[th]
	\centerline{\includegraphics[scale=1]{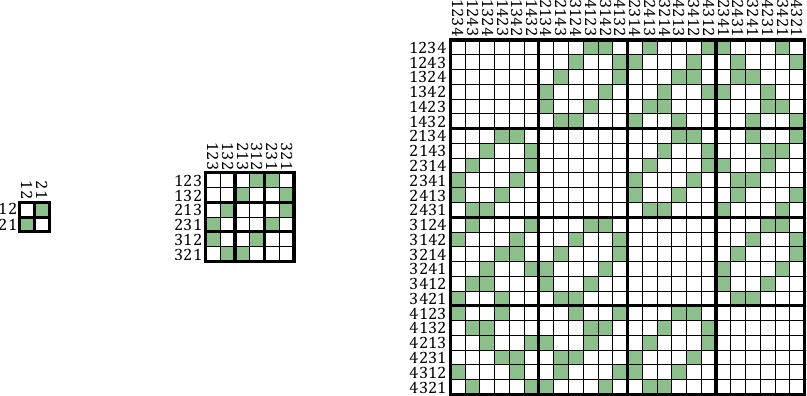}}
	\caption{The matrices $P^{(2)}$, $P^{(3)}$ and $P^{(4)}$.}
	\label{f:matrix_for_a_permutation_2_3_4}
\end{figure}

The form of the matrix $P^{(k)}$ for $k=2,3,4$
is presented in Figure~\ref{f:matrix_for_a_permutation_2_3_4}.
White squares represent zeroes, the rest of the squares contain 1.

As proved in the earlier paper,
a certain submatrix of the communication matrix for $n$-state 2DFA
can be linearly transformed
into a block-diagonal matrix
of order $\sum_{k=1}^n {n \choose{k-1}} {n \choose k} k!$,
which consists of
$\binom{n}{k-1} \binom{n}{k}$ copies of each submatrix $P^{(k)}$~\cite[Lemmata 4, 11, 13]{PetrovOkhotin_2dfa_to_1ufa}.
At the same time, it is shown the entire comminication matrix for $n$-state 2DFA
has the same rank as the latter block-diagonal matrix~\cite[Thm.~3]{PetrovOkhotin_2dfa_to_1ufa}.
Overall, the rank of the communication matrix is given by the formula in the next theorem.

\begin{oldtheorem}[\cite{PetrovOkhotin_2dfa_to_1ufa}]\label{lower_bound_rank_conversion_theorem}
For every $n \geqslant 1$,
the rank of the communication matrix for $n$-state 2DFA is exactly
$\sum_{k=1}^n {n \choose{k-1}} {n \choose k} \rank P^{(k)}$.
\end{oldtheorem}

Hence, in order to determine the rank of the communication matrix,
it is left to obtain the rank of $P^{(k)}$, for each $k$.
For convenience, rows of $P^{(k)}$ shall be permuted by indexing them with inverses of permutations.
Formally, $P^{(k)}$ is replaced
with another $k! \times k!$ matrix $Q^{(k)}$,
defined by $Q^{(k)}_{\pi, \sigma} = P^{(k)}_{\pi^{-1}, \sigma}$,
for every pair of permutations $\pi$ and $\sigma$;
that is, $Q^{(k)}_{\pi, \sigma} = 1$ if $\sigma \circ \pi^{-1}$ is cyclic.
The rank of $Q^{(k)}$ is the same as the rank of $P^{(k)}$,
and the goal is to determine it.

\section{The matrix as an action on group algebra}\label{section_group_algebra}

In order to determine the rank of the matrix $Q^{(n)}$, with $n \geqslant 1$,
this matrix shall be represented as an action on a group algebra.

Let $S_n$ be the group of permutations on $n$ elements.
Consider the \textit{group algebra} $\mathbb{C}[S_n]$:
this is a linear space of dimension $n!$ over the complex numbers,
and with the basic vectors identified with elements of $S_n$,
endowed by a multiplication operation.
The addition operation is as per the definition of a linear space,
\begin{align*}
	\Big( \sum_{p \in S_n} a_p p \Big)
	+
	\Big( \sum_{q \in S_n} b_q q \Big)
		&=
	\sum_{p \in S_n} (a_p + b_p) p
\intertext{%
and the multiplication operation is extended from $S_n$ as follows:
}
	\Big( \sum_{p \in S_n} a_p p \Big)
	\Big( \sum_{q \in S_n} b_q q \Big)
		&=
	\sum_{p, q \in S_n} a_p b_q (p \circ q).
\end{align*}

Let $C_n \subseteq S_n$ be the set of all $(n-1)!$ cyclic permutations,
and denote their sum
by $$q_n = \sum_{p \in C_n} p\in \mathbb{C}[S_n].$$

\begin{lemma}\label{q_n_is_central}
The sum of the cyclic permutation, $q_n$,
belongs to the center of algebra $\mathbb{C}[S_n]$,
that is, $xq_n=q_nx$ for every $x\in \mathbb{C}[S_n]$.
\end{lemma}

\begin{proof}
By linearity, it suffices to consider the case $x\in S_n$.
Using the known fact that for every permutation $p \in S_n$,
the conjugate permutation $x \circ p \circ x^{-1}$ has the same lengths of cycles,
if $p$ is cyclic, then so is $x \circ p \circ x^{-1}$.
Furthermore, since the map $p \mapsto x \circ p \circ x^{-1}$ is bijective on $S_n$,
it is bijective on $C_n$. 
Therefore,
$$
	xq_n
		=
	\sum_{p\in C_n} x \circ p
		=
	\sum_{p\in C_n} (x \circ p \circ x^{-1} \circ x)
		=
	\Big(\sum_{p\in C_n} (x \circ p \circ x^{-1})\Big)x
		=
	\Big(\sum_{q\in C_n} q\Big)x
		=
	q_nx.
$$
\end{proof}

Consider the linear operator $\mathcal{L}_n$ on the space $\mathbb{C}[S_n]$ defined 
by $x \mapsto q_nx$. Look at the matrix of this operator in the standard basis of $\mathbb{C}[S_n]$
(identified with $S_n$).

\begin{lemma}\label{Q_is_long_cycle_multiplication_lemma}
The matrix of $\mathcal{L}_n$ in the standard basis is the matrix $Q^{(n)}$.
\end{lemma}
\begin{proof}
For every $g\in S_n$, it has to be proved that
$\sum_{h \in S_n} Q^{(n)}_{g,h} h = \mathcal{L}_n(g)$.

By the definition of $Q^{(n)}$,
the coefficient $Q^{(n)}_{g,h}$ at $h$ is $1$ if $h \circ g^{-1} \in C_n$,
and $0$ otherwise.
Then it can be transformed as follows,
where a substitution of $p$ for $h \circ g^{-1}$ is made at the last step.
$$
	\sum_{h \in S_n} Q^{(n)}_{g,h} h
		=
	\sum_{h\in S_n: h \circ g^{-1} \in C_n} h
		=
	\sum_{h\in S_n: h \circ g^{-1} \in C_n} h \circ g^{-1} \circ g
		=
	\sum_{p\in C_n} p \circ g
		=
	\mathcal{L}_n(g)
$$
\end{proof}

So, the rank of matrix $Q^{(n)}$ is equal to the rank of operator $\mathcal{L}_n$,
with which we shall work from now on.

\section{Representations and modules}\label{section_modules}

The calculation of the rank of $\mathcal{L}_n$
is based on the representation theory of the symmetric group $G=S_n$.

Here we briefly recall the necessary definitions and facts about the representations of finite groups.
In this section, $G$ is a finite group (but further we need only the case $G=S_n$).

\begin{definition}
A \emph{$G$-module}
is a vector space $V$ over the complex numbers,
with a homomorphism $\varphi \colon G \to GL(V)$,
where $GL(V)$ is the set of all invertible linear mappings $V \to V$.
The function $\varphi$ then describes
the action of the elements of $G$ on the elements of $V$:
$gv$ for $g\in G$, $v\in V$ is defined as $\varphi(g)(v)$. The homomorphism property
yields the associativity $(gh)v=g(hv)$ for $g,h\in G$, $v\in V$. 
\end{definition}

$G$-modules are often called \emph{representations} of group $G$
(strictly speaking, a \emph{representation} is a map $\varphi$).

\begin{definition}
The \emph{dimension} of a $G$-module $(V, \varphi)$ is the dimension of a vector space $V$.
\end{definition}

For a $G$-module $(V, \varphi)$, the function $\varphi$
is extended by linearity to a homomorphism on the group algebra $\mathbb{C}[G]$.
This extended homomorphism maps elements of the group algebra $\mathbb{C}[G]$
to linear mappings on $V$ that are not necessarily invertible
(that is, to the algebra $\End(V)$ of linear endomorphisms of $V$).
$$
	\varphi \big( \sum_{g \in G} a_g g \big) = \sum_{g \in G} a_g \varphi(g)
$$

\begin{definition}
A $G$-module $(V, \varphi)$ is called \emph{reducible},
if there exists a non-trivial proper subspace $W \subset V$
invariant to the action of elements of $G$.
Otherwise, the module is \emph{irreducible}.
\end{definition}

\begin{oldtheorem}[{Maschke, see Sagan~\cite[Thm.~1.5.3]{Sagan}}] \label{maschke_theorem}
Let $G$ be a finite group, and let $(V, \varphi)$ be a finite-dimensional $G$-module.
Then there is a number $d$ 
and irreducible $G$-modules $W^{(1)}, \dots, W^{(d)}$,
such that $V$ is a direct sum of the $W^{(i)}$:
$V = W^{(1)} \oplus \dots \oplus W^{(d)}$.
\end{oldtheorem}

In this direct sum representation, each of the linear operators $\varphi(x)$ for $x\in \mathbb{C}[G]$
is invariant on each of the irreducible $G$-modules $W^{(i)}$.
Thus, for the operator
$\varphi(x)$ we have
\begin{equation}\label{rank_formula_1}
\rank \varphi(x)=\sum_{i=1}^d \rank(\varphi(x)|_{W^{(i)}}).
\end{equation}

We use the following version of Schur's lemma.

\begin{lemma}[Schur]\label{Schur}
Let $G$ be a finite group, and let $(V, \varphi)$ be a finite-dimensional $G$-module.
Let $y\in \mathbb{C}[G]$.
Then every irreducible submodule
$W\subset V$ is $y$-invariant
(that is, $\varphi(y)$ maps $W$ to $W$),
and, furthermore, if $y$ belongs to the center of algebra $\mathbb{C}[G]$,
then the restriction of $\varphi(y)$ to $W$ is a scalar operator. 
\end{lemma}

\begin{proof}
First, since $W$ is a $G$-module,
it is invariant under action of each $\varphi(g)$, with $g\in G$.
Since $\varphi(y)$ is a linear combination of $\varphi(g)$'s,
the subspace $W$ is also invariant under action of $\varphi(y)$.

Turning to the second part, let $y$ belong to the center of $\mathbb{C}[G]$,
and let $a$ be any eigenvalue of the restriction $y|_W$ of $\varphi(y)$ to $W$,
that is, $\varphi(y)(w)=aw$ for some vector $w \in W$.
Replacing $y$ with $y'=y-a\cdot e$ (where $e$ is the unit element of the group $G$),
we obtain $\varphi(y')(w)=\varphi(y-a \cdot e)(w)=aw-aw=0$.
This $y'$ is also a central element (as a sum of two central elements).
Then the kernel $W_1$ of $y'|_W$ is non-trivial (as it contains $w$).
But it is an invariant subspace of every $g\in G$:
for $v\in W_1$, $y'(gv)=(y'g)v=(gy')v=g(y'v)=g(0)=0$, thus $gv\in W_1$.
Since $W$ is irreducible, we get $W_1=W$,
so $y' \cdot w=0$ for all $w \in W$.
Then, $(y-a \cdot e) \cdot w = 0$
and hence $yw = aw$ for all $w \in W$,
that is, $y$ is indeed scalar on $W$.
\end{proof}

Consider a $G$-module defined by the vector space $\mathbb{C}[G]$,
with the homomorphism $\varphi$
mapping a permutation $g$ to a linear operator on $\mathbb{C}[G]$
which is a left multiplication by $g$.
By Maschke's theorem for this $G$-module,
the group algebra $\mathbb{C}[G]$
is decomposed into a direct sum of irreducible $G$-modules.

In particular, if $G=S_n$, then $\mathcal{L}_n=\varphi(q_n)$ by definition,
and the rank of the operator $\mathcal{L}_n$
is the sum of the ranks of $\varphi(q_n)$ on the irreducible $S_n$-modules
in the representation of $\mathbb{C}[S_n]$ according to Maschke's theorem.
The question is, how many irreducible $S_n$-modules are there in this decomposition,
and how many times each of them occurs?
The next theorem provides an answer.

Recall that two elements $g,h\in G$ are called \emph{conjugate}, if $g=xhx^{-1}$ for some $x\in G$. 
Conjugacy is an equivalence relation, with classes called \emph{conjugacy classes}.

\begin{oldtheorem}[{\cite[Prop.~1.10.1]{Sagan}}]\label{group_algebra_representation_theorem}
Up to isomorphism, there are finitely many irreducible $G$-modules,
and they correspond to the conjugacy classes of the group $G$.

Every irreducible $G$-module $(V, \varphi)$
occurs in the decomposition of the group algebra $\mathbb{C}[G]$ into irreducible $G$-modules exactly
$\dim V$ times.
\end{oldtheorem}

For isomorphic modules $W^{(i)}$, $W^{(j)}$,
the restrictions of $\varphi(x)$ to these moduli
are isomorphic operators, in particular, they have equal rank.
Therefore, in view of Theorem \ref{group_algebra_representation_theorem},
\eqref{rank_formula_1} for the operator $\mathcal{L}_n$ reads as

\begin{equation}\label{rank_formula}
	\rank \mathcal{L}_n
		=
	\sum_{(V, \varphi)} \dim V \cdot \rank(\mathcal{L}_n|_V),
\end{equation}
where the sum is over all irreducible representations $(V, \varphi)$
of the symmetric group $S_n$.
By Lemma \ref{Schur}, the restrictions $\mathcal{L}_n|_V$ are scalar operators,
and each of them has rank $\dim V$
(unless it is a zero operator, that is, a multiplication by zero).
Therefore,

\begin{equation}\label{rank_formula_better}
\rank \mathcal{L}_n=\sum_{(V, \varphi): \mathcal{L}_n|_V \ne 0} (\dim V)^2.
\end{equation}

It remains to determine for which irreducible $S_n$-modules the restrictions of $\mathcal{L}_n$
onto $(V, \varphi)$ are non-zero.
Then the desired rank is the sum of squares of dimensions.
This is done in the next section.

\section{Specht modules}\label{section_specht}

The form of irreducible $S_n$-modules is well-studied:
they are indexed by partitions of $n$,
and are known as \emph{Specht modules}, see Sagan~\cite[Ch.~2]{Sagan}.

\begin{figure}[t]
\center{\includegraphics[width=0.6\linewidth]{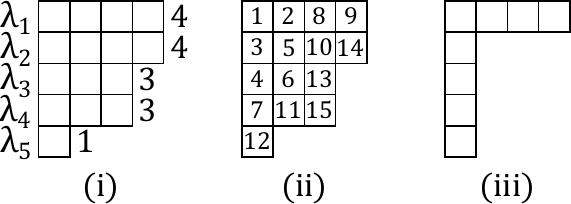}}
\caption{(i) Young diagram for $\lambda=(4, 4, 3, 3, 1)$;
	(ii) a standard Young tableau;
	(iii) a Young diagram of a hook shape, with $\lambda=(4, 1, 1, 1, 1)$.
	}
\label{f:young_diagrams}
\end{figure}

Let $n \geqslant 1$ be an integer,
and let $\lambda = (\lambda_1, \dots, \lambda_m)$
be a partition of $n$, with $n=\lambda_1 + \ldots + \lambda_m$
and $\lambda_1 \geqslant \ldots \geqslant \lambda_m \geqslant 1$.
Every partition has a corresponding \emph{Young diagram} $Y(\lambda)$,
as in Figure~\ref{f:young_diagrams}(i),
which consists of $m$ rows,
with each $i$-th row of length $\lambda_i$.

\begin{definition}
A \emph{standard Young tableau} of size $n$
is a Young diagram,
with its boxes filled with the elements $\{1, \ldots, n\}$,
with each element occurring exactly once, and the numbers in every row and the numbers in every column 
strictly increasing.
\end{definition}

A standard Young tableau is illustrated in Figure~\ref{f:young_diagrams}(ii).

The conjugacy classes in $S_n$ are parametrized by partitions
(this follows from the known fact that two permutations are conjugate if and only if
they have the same lengths of cycles),
thus, by Theorem \ref{group_algebra_representation_theorem},
the irreducible $S_n$-modules may be also parametrized by partitions.
There is a known canonical way
to do this which is known as \emph{Specht modules} $W^\lambda$, for a partition $\lambda$ of $n$.
We do not describe them here, but use the following fact about them.
\begin{oldtheorem}[{\cite[Thm.~2.6.5]{Sagan}}]\label{Specht_module_dimension_theorem}
The dimension of $W^\lambda$ equals the number of standard Young tableaux
of shape $\lambda$.
\end{oldtheorem}

For finding for which $\lambda$ the action of $\mathcal{L}_n$ on $W^\lambda$ is non-zero, we need
a piece of theory of \emph{characters} of $S_n$.

\section{Characters and the Murnaghan--Nakayama rule}\label{section_characters}

\begin{definition}
Let $G$ be a group, and let $(V,\varphi)$ be a $G$-module.
A \emph{character} of $(V,\varphi)$ is a function
$\chi_{\varphi}(x) = \mathop{\mathrm{tr}} \varphi(x)$ for $x\in \mathbb{C}[G]$.
\end{definition}

Note that if $x\in \mathbb{C}[G]$ is central and $(V, \varphi)$ is an irreducible $G$-module,
the action of $x$ on $V$ is non-zero
if and only if the character of $(V, \varphi)$ at $x$ is non-zero.
Indeed, the action of $x$ on $V$ is multiplication by a constant:
$\varphi(x)(v)=av$ for $v \in V$.
Then the trace of $\varphi(x)$ is $a \cdot \dim V$.
Since $\dim V \neq 0$, the trace is zero if and only if $a=0$.

It is known that the function $\chi$ is constant on each conjugacy class of $G$,
which is stated in the next theorem.

\begin{oldtheorem}[{\cite[Prop.~1.8.5, pt. 2]{Sagan}}] \label{conjugacy_class_equality_theorem}
Let $G$ be a group, and let $(V,\varphi)$ be a $G$-module with character $\chi$.
Let $K$ be a conjugacy class of $G$, and let $g, h \in K$.
Then, $\chi(g) = \chi(h)$.
\end{oldtheorem}

Since the class $C_n$ of cyclic permutations forms a conjugacy class,
$q_n=\sum_{p \in C_n} p$ is a sum of $p$ over this class.
Then, for $S_n$-module, the value of its character $\chi$ on $q_n$
can be expressed through its one value on a single arbitrary cyclic permutation $\widehat{p}$,
as $\chi(q_n)=\sum_{p \in C_n} \chi(p)=|C_n| \cdot \chi(\widehat{p}) = (n-1)! \cdot \chi(\widehat{p})$.
Hence, $\chi(q_n) \neq 0$ if and only if $\chi(\widehat{p}) \neq 0$.

Putting all of this together,
and denoting by $\chi_\lambda$
the character of the Specht module $W^\lambda$ corresponding to a partition $\lambda$ of $n$,
the expression for the rank~\eqref{rank_formula_better} reads as
\begin{equation}\label{rank_even_better}
	\rank \mathcal{L}_n
		=
	\sum_{\lambda: \mathcal{L}_n|_{W^\lambda}\ne 0} (\dim W^\lambda)^2
		=
	\sum_{\lambda: \chi_\lambda(q_n)\ne 0} (\dim W^\lambda)^2
		=
	\sum_{\lambda: \chi_\lambda(\widehat{p})\ne 0} (\dim W^\lambda)^2,
\end{equation}
where the summation is taken over all partitions $\lambda$ of $n$.

Therefore, the question of finding the rank of $\mathcal{L}_n$
is reduced to determining which partitions $\lambda$
have their characters $\chi_\lambda$ non-zero at a cyclic permutation $\widehat{p}$.
There is a rule that helps to answer this question.

\begin{definition}
Let $\lambda = (\lambda_1, \dots, \lambda_m)$ be a partition of $n$.
A \emph{skew hook}, or \emph{rim hook}, of $\lambda$ is a diagram $\xi$
obtained by taking all cells on a finite lattice path
with steps one unit northward or eastward,
such that the entire path lies inside a Young diagram $Y_\lambda$,
and $Y_\lambda \setminus \xi$ is also a Young diagram.
\end{definition}

\begin{figure}[t]
\center{\includegraphics[scale=1]{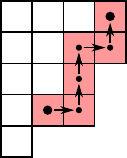}}
\caption{An example of a skew hook.}
\label{f:skew_hook}
\end{figure}

An example of a skew hook is presented in Figure~\ref{f:skew_hook}.
The cells that are included in a skew hook are highlighted.

Note that a skew hook cannot contain a $2 \times 2$ square of cells, because each step on the path increases the sum of cell coordinates, and a $2 \times 2$ square contains two cells with equal sum of coordinates.

\begin{definition}
Let $\xi$ be a skew hook. The \emph{leg length} of $\xi$ is denoted $ll(\xi)$ and is equal to the number of rows in $\xi$ minus 1.
\end{definition}

Denote by $\lambda \setminus \xi$ the partition
which corresponds to a Young diagram $Y_\lambda \setminus \xi$.

\begin{oldtheorem}[Murnaghan--Nakayama rule, see Sagan {\cite[Thm.~4.10.2]{Sagan}}]\label{murnaghan_nakayama_rule}
Let $\lambda$ be a partition of $n$, and let $\alpha$ be a permutation with cycle lengths
$\alpha_1\geqslant \alpha_2\geqslant \cdots \geqslant \alpha_k>0$,
$\sum_{i=1}^k \alpha_i = n$.
Let $\alpha'$ be a permutation of $n-\alpha_1$ elements
with cycle lengths $\alpha_2, \alpha_3, \dots, \alpha_n$.
Then, $\chi_\lambda(\alpha) = \sum_{\xi} (-1)^{ll(\xi)} \chi_{\lambda \setminus \xi}(\alpha')$,
where the sum is over all skew hooks $\xi$ of $\lambda$ having $\alpha_1$ cells.
\end{oldtheorem}

Of particular interest is the case $k = 1$,
that is, when $\alpha$ is a cyclic permutation.
Then, $\alpha'$ is an empty partition and the only element of $S_0$,
with $\mathbb{C}[S_0]$ being irreducible.
The character in this case is $\chi_{\lambda \setminus \xi}(\alpha') = \dim \mathbb{C}[S_0] = 1$.

With this rule, the question of which characters are non-zero can be answered precisely.

\begin{definition}
A partition $\lambda$ of $n$
is called a \emph{hook-shaped} partition if $\lambda_i = 1$ for $i \geq 2$.
\end{definition}

\begin{lemma} \label{hook_supremacy_lemma}
Let $\lambda$ be a partition of $n$, let $\widehat{p}$ be a cyclic permutation. 
Then, $\chi_\lambda(\widehat{p})$ is non-zero if and only if $\lambda$ is hook-shaped.
\end{lemma}

\begin{proof}
By the Murnaghan--Nakayama rule for a cyclic permutation,
$\chi_\lambda(\widehat{p}) = \sum_{\xi} (-1)^{ll(\xi)} \chi_{\lambda \setminus \xi}\big(()\big)$,
where the character $\chi_{\lambda \setminus \xi}$ is evaluated on the empty permutation $() \in S_0$,
and the sum is over all skew hooks $\xi$ of $\lambda$ having $n$ cells.

Since $\xi$ has $n$ cells, $Y(\lambda)$ also has $n$ cells, and $\xi$ is a subset of $Y(\lambda)$,
they must be equal.

Note that $\lambda$ is hook-shaped if and only if the respective Young diagram $Y(\lambda)$ does not contain a $2 \times 2$ block of cells.
Therefore, if $\lambda$ is not hook-shaped, there are no valid $\xi$, and the sum is zero.

On the other hand, if $\lambda$ is hook-shaped, there is exactly one $\xi$ satisfying the requirements (namely, $Y(\lambda)$). 
Since $(-1)^{ll(\xi)}$ and $\chi_{\lambda \setminus \xi}\big(()\big) = 1$
are both non-zero,
their product $\chi_\lambda(\widehat{p})$ is also non-zero in this case.
\end{proof}

\begin{theorem}\label{rank_of_P_complex_theorem}
$\rank Q^{(n)}={2n-2\choose n-1}$.
\end{theorem}
\begin{proof}
Using Lemma \ref{hook_supremacy_lemma}, the above expression~\eqref{rank_even_better} reads as
\begin{equation*}
	\rank Q^{(n)}
		=
	\rank \mathcal{L}_n
		=
	\sum_{\lambda: \chi_\lambda(\widehat{p})\ne 0} (\dim W^\lambda)^2
		=
	\sum_{\lambda \, \text{is hook-shaped}} (\dim W^\lambda)^2.
\end{equation*}
A hook-shaped diagram $\lambda$ consists of the row of length $k$ and a column of length $n+1-k$. 
Clearly, the number of $\lambda$-shaped standard Young tableaux equals ${n-1\choose k-1}$:
we fix which elements of $\{2,3,\ldots,n\}$ belong to the first row,
arrange them in increasing order, and arrange remaining elements
in the increasing order in the first column.
Then, by Theorem~\ref{Specht_module_dimension_theorem},
the dimension of $W^\lambda$ is ${n-1\choose k-1}$.
Therefore,
\begin{equation*}
	\rank \mathcal{L}_n
		=
	\sum_{\lambda \, \text{is hook-shaped}} (\dim W^\lambda)^2
		=
	\sum_{k=1}^{n} {n-1\choose k-1}^2
		=
	\sum_{k=1}^{n} {n-1\choose k-1}{n-1\choose n-k}
\end{equation*}
The latter sum gives the number of ways to choose $k-1$ elements among $\{1,2,\ldots,n-1\}$
and $n-k$ elements among $\{n,n+1,\ldots, 2n-2\}$, for all $k$.
This is the same as choosing $n-1$ elements among $\{1,2,\ldots,2n-2\}$,
which is $\binom{2n-2}{n-1}$.
\end{proof}

\section{Asymptotics of the rank of the communication matrix}\label{section_asymptotics}

\begin{theorem}\label{communication_matrix_theorem}
For every $n$, the rank of the communication matrix for $n$-state 2DFA
is exactly
$\sum_{k=1}^n {n \choose k - 1}{n \choose k}{2k - 2 \choose k - 1}$.
\end{theorem}
\begin{proof}
By Theorem~\ref{lower_bound_rank_conversion_theorem}, for every $n$,
the rank of the communication matrix $n$-state 2DFA is
$\sum_{k=1}^n {n \choose{k-1}} {n \choose k} \rank P^{(k)}$.
By Theorem~\ref{rank_of_P_complex_theorem},
$\rank P^{(k)} = {2k - 2 \choose k - 1}$.
\end{proof}

\begin{theorem}\label{asymptotics_theorem}
The numbers
$\boundfunc{n}=\sum_{k=1}^n {n \choose k - 1}{n \choose k}{2k - 2 \choose k - 1}$
are of the order 
$(1+o(1)) \frac{3 \sqrt{3}}{8 \pi n} \cdot 9^n$.
\end{theorem}
\begin{proof}
We have
$$
	9^{-n}\boundfunc{n}=\sum_{k=1}^n a_k,
$$
where
\begin{equation*}
	a_k
		=
	\frac12
	{n\choose k-1}\left(\frac23\right)^{k-1}\left(\frac13\right)^{n-k+1}
	\cdot
	{n\choose k}\left(\frac23\right)^{k}\left(\frac13\right)^{n-k}
	\cdot
	2^{2-2k}{2k-2\choose k-1}.
\end{equation*}
The idea is to use the de Moivre--Laplace local central limit theorem,
which ensures that
\begin{equation*}
	{n\choose t}p^t(1-p)^{n-t}
		\sim
	\frac{1}{\sqrt{2\pi np(1-p)}}
	e^{-\frac{(t-pn)^2}{2p(1-p)n}}
\end{equation*}
uniformly in the range $|t-np|<n^{3/5}$.
This theorem is applied three times,
that is, to each of the three factors in the above formula for $a_k$.
For the first factor, it is used for $p=\frac{2}{3}$ and $t=k-1$:
\begin{equation*}
	{n\choose k-1}\left(\frac23\right)^{k-1}\left(\frac13\right)^{n-k+1}
		\sim
	\frac{1}{\sqrt{2\pi n \cdot 2/9}}
	e^{-\frac{(k-n \cdot 2/3)^2}{4/9 \cdot n}}
	\quad (\text{for } |k-1-n \cdot 2/3|<n^{3/5})
\end{equation*}
For the second factor, $p=\frac{2}{3}$ and $t=k$ give the same approximation:
\begin{equation*}
	{n\choose k}\left(\frac23\right)^{k}\left(\frac13\right)^{n-k}
		\sim
	\frac{1}{\sqrt{2\pi n \cdot 2/9}}
	e^{-\frac{(k-n \cdot 2/3)^2}{4/9 \cdot n}}
	\quad (\text{for } |k-n \cdot 2/3|<n^{3/5})
\end{equation*}
Applying the de Moivre--Laplace theorem for the third factor
requires $p=1/2$, $t=k-1$ and $2k-2$ instead of $n$.
\begin{multline*}
	{2k-2 \choose k-1}\left(\frac12\right)^{k-1}\left(\frac12\right)^{k-1}
		\sim
	\frac{1}{\sqrt{2\pi (2k-2) \cdot 1/4}}
	e^{-\frac{(k-1-1/2 \cdot (2k-2))^2}{1/2 \cdot (2k-2)}}
		=
	\frac{1}{\sqrt{\pi (k-1)}}
	e^{-\frac{0^2}{k-1}}
		= \\ =
	\frac{1}{\sqrt{\pi (k-1)}}
	\qquad (\text{for } |k-1-(2k-2) \cdot 1/2|<(2k-2)^{3/5})
\end{multline*}
The latter estimation is actually valid for all $k$ (since $k-1-(2k-2) \cdot 1/2$ is equal to 0),
but is used for $|k-2n/3|<n^{3/5}$.
In this case, for $n$ approaching infinity, $k \sim 2n/3$, and therefore,
\begin{equation*}
	{2k-2 \choose k-1}\left(\frac12\right)^{k-1}\left(\frac12\right)^{k-1}
		\sim
	\frac{1}{\sqrt{\pi (k-1)}}
		\sim
	\frac{1}{\sqrt{\pi n \cdot 2/3}}
\end{equation*}
Overall, we have
\begin{equation*}
	a_k
		\sim
	\frac{1}{2}
	\cdot
	\frac{1}{2\pi n\cdot 2/9}\cdot \frac{1}{\sqrt{\pi\cdot 2n/3}}
	\cdot
	e^{-\frac{(k-2n/3)^2}{n\cdot 2/9}}.
\end{equation*}
Using the integral 
approximation of the sum
and then substituting $t=x/\sqrt{n \cdot 2/9}$
yields
\begin{equation*}
	\sum_{|k-2n/3|<n^{3/5}} 
	e^{-\frac{(k-2n/3)^2}{n\cdot 2/9}}
		\sim
	\int_{-n^{3/5}}^{n^{3/5}} 
	e^{-\frac{x^2}{n\cdot 2/9}}dx
		=
	\sqrt{n\cdot 2/9}\int_{-n^{1/10}/\sqrt{2/9}}^{n^{1/10}/\sqrt{2/9}}
	e^{-t^2}dt
	\sim
	\sqrt{\pi n\cdot 2/9}
\end{equation*}
Finally, we get 
\begin{equation*}
	\sum_{|k-2n/3|<n^{3/5}} a_k
		\sim
	\frac{1}{2}
	\cdot
	\frac{1}{2\pi n\cdot 2/9}\cdot \frac{1}{\sqrt{\pi\cdot 2n/3}}
	\cdot
	\sqrt{\pi n\cdot 2/9}
		=
	\frac{3\sqrt{3}}{8\pi n}
\end{equation*}

As for the range $|k-2n/3|>n^{3/5}$,
we may use that all three factors
$$
{n\choose k-1}\left(\frac23\right)^{k-1}\left(\frac13\right)^{n-k+1};\,
{n\choose k}\left(\frac23\right)^{k}\left(\frac13\right)^{n-k};\,
2^{2-2k}{2k-2\choose k-1}
$$
in the formula for $a_k$
are certain binomial probabilities
and hence do not exceed 1.
In the following upper bound,
the first and the third factors are replaced by 1,
while the second factor is represented as the probability
for a binomially distributed random variable $X$ with parameters $n$ and $\frac{2}{3}$
to be exactly $k$.
\begin{multline*}
	\sum_{|k-2n/3| \geqslant n^{3/5}}
	a_k
		\leqslant
	\frac12\sum_{|k-2n/3| \geqslant n^{3/5}}
	{n\choose k}\left(\frac23\right)^{k}\left(\frac13\right)^{n-k}
		=
	\frac12\sum_{|k-2n/3| \geqslant n^{3/5}}
	\mathrm{Pr}(X=k)
		= \\ =
	\frac12 \mathrm{Pr}\big(|X-2n/3| \geqslant n^{3/5}\big)
\end{multline*}
This probability
is super-polynomially small by the central limit
theorem, thus this part of the sum of $a_k$'s
does not influence the asymptotics,
and we get
$$
	\boundfunc{n}
		=
	\Big(
	\sum_{|k-2n/3|<n^{3/5}} a_k
	+
	\sum_{|k-2n/3| \geqslant n^{3/5}} a_k
	\Big)
	\cdot 9^n
		=
	\bigg(
	\frac{3\sqrt{3}}{8\pi n} + o\Big(\frac{1}{n}\Big)
	\bigg)\cdot 9^n
		\sim
	\frac{3\sqrt{3}}{8\pi n}
	9^n
$$
\end{proof}

The values of the lower bound established in Theorem~\ref{lower_bound_asymptotics_theorem}
for small values of $n$
are presented in Table~\ref{tab:bounds_for_small_N}.
For $n$ up to 3, the lower bound matches the upper bound,
and then it starts to diverge.

Now, by Schmidt's theorem (Theorem~\ref{Schmidt_theorem}),
the rank of the communication matrix for $n$-state 2DFA
determined in Theorem~\ref{communication_matrix_theorem}.
is a lower bound on the number of states needed to transform an $n$-state 2DFA to a UFA.

\begin{corollary}\label{lower_bound_asymptotics_theorem}
For every $n$, there is a language recognized by a 2DFA with $n$ states,
such that every UFA for the same language
requires at least
$\sum_{k=1}^n {n \choose k - 1}{n \choose k}{2k - 2 \choose k - 1}$
states,
which is asymptotically of the order $\frac{3\sqrt{3}}{8\pi n} 9^n$.
\end{corollary}

\begin{table}[h]
\caption{Bounds on the complexity of transforming $n$-state 2DFA to UFA, for small values of $n$.}
\label{tab:bounds_for_small_N}
\begin{center}
\begin{tabular}{|c|r|r|r|}
\hline
\multirow{2}{*}{$n$}
& earlier lower bound
& \textbf{new lower bound}
& upper bound \\
\cline{2-4}
& $\sum_{k=1}^n {n \choose{k-1}} {n \choose k} 2^{k-1}$
& $\sum_{k=1}^n \binom{n}{k-1} \binom{n}{k} \binom{2k-2}{k-1}$
& $\sum_{k=1}^n \binom{n}{k-1} \binom{n}{k} k!$ \\
\hline
1 & 1 & 1 & 1 \\
\hline
2 & 6 & 6 & 6 \\
\hline
3 & 33 & 39 & 39 \\
\hline
4 & 180 & 276 & 292 \\
\hline
5 & 985 & 2\,055 & 2\,505 \\
\hline
6 & 5\,418 & 15\,798 & 24\,306 \\
\hline
7 & 29\,953 & 124\,173 & 263\,431 \\
\hline
8 & 166\,344 & 992\,232 & 3\,154\,824 \\
\hline
9 & 927\,441 & 8\,030\,943 & 41\,368\,977 \\
\hline
10 & 5\,188\,590 & 65\,672\,850 & 589\,410\,910 \\
\hline
\end{tabular}
\end{center}
\end{table}

\section{Conclusion and future work}

With the rank of the communication matrix determined precisely,
the lower bound on the state complexity of transforming 2DFA to UFA
is accordingly improved, as shown in Table~\ref{tab:bounds_for_small_N}.
This bound is the best that could be achieved
by the method of Schmidt based on ranks of matrices~\cite[Thm.~3]{PetrovOkhotin_2dfa_to_1ufa}.
Whether this bound is optimal, is not known,
but the known upper bound on the required number of states,
given in the last column of Table~\ref{tab:bounds_for_small_N},
is of a higher order of magnitude.

It might be possible to improve this lower bound
by estimating the \emph{positive integer rank} of the communication matrix for $n$-state 2DFA,
where positive integer rank is defined as the least number of vectors of nonnegative integers,
such that every row of the matrix is a linear combination of these vectors
with nonnegative integer coefficients~\cite[Lem.~2.1]{CohenRothblum}.
In fact, Schmidt's theorem uses the rank of the communication matrix
to approximate its positive integer rank;
no other methods for proving lower bounds on the positive integer rank are known,
and calculating it is a computationally hard problem~\cite{Vavasis}.

\section*{Acknowledgement}

This work was supported by the Russian Science Foundation, project 23-11-00133.

\end{document}